\documentclass[preprint]{aastex}
\usepackage{amsmath}
\newcommand{\beq}{\begin{equation}}
\newcommand{\eeq}{\end{equation}}
\newcommand{\bea}{\begin{eqnarray}}
\newcommand{\eea}{\end{eqnarray}}
\newcommand{\rem}[1]{ }

\begin{document}
\title{On Poynting-Flux-Driven Bubbles and Shocks Around Merging Neutron Star Binaries}

\author{Mikhail V. Medvedev\altaffilmark{1}} 
\affil{Department of Physics and Astronomy, University of Kansas, Lawrence, KS 66045}
\altaffiltext{1}{Also at the ITP, NRC ``Kurchatov Institute", Moscow 123182, Russia}

\author{Abraham Loeb} 
\affil{Astronomy Department, Harvard University, 60 Garden St., Cambridge, MA 02138}

\begin{abstract}
Merging binaries of compact relativistic objects (neutron stars and black holes) are thought to be progenitors of short gamma-ray bursts and sources of gravitational waves, hence their study is of great importance for astrophysics. Because of the strong magnetic field of one or both binary members and high orbital frequencies, these binaries are strong sources of energy in the form of Poynting flux (e.g., magnetic-field-dominated outflows, relativistic leptonic winds, electromagnetic and plasma waves). The steady injection of energy by the binary forms a bubble (or a cavity) filled with matter with the relativistic equation of state, which pushes on the surrounding plasma and can drive a shock wave in it. Unlike the Sedov-von Neumann-Taylor blast wave solution for a point-like explosion, the shock wave here is continuously driven by the ever-increasing pressure inside the bubble. We calculate from the first principles the dynamics and evolution of the bubble and the shock surrounding it and predict that such systems can be observed as radio sources a few hours before and after the merger. At much later times, the shock is expected to settle onto the Sedov-von Neumann-Taylor solution, thus resembling an explosion.  
\end{abstract}

\section{Introduction}

Neutron star-neutron star (NS-NS) and neutron star-black hole (NS-BH) binaries are of great interest in astrophysics. First, they are thought to be progenitors of short gamma-ray bursts (GRBs), though the evidence is based mostly on the shortness of the time-scale of the final merger process, in addition to the energetics considerations and population studies \citep{Blinnikov+84,Eichler+89,Nakar07,Berger11}. Hence, the natural question arises: Are there any observational signatures that can help to tell apart the merging binary progenitor from alternative GRB models? Unlike other short GRB studies (including those studying their radio emission, e.g., \citealp{PP10,NakarPiran11}), we investigate here what happens to the system long before the merger, not during or after. Second, merging binaries are also considered to be strong sources of gravitational waves, which can potentially be observed with upcoming detectors (e.g., Advanced LIGO). Cross-correlating the gravitational wave signal with astrophysical objects observed using other techniques will advance astrophysics drastically. 

\begin{figure}[t]
\center
\includegraphics[angle = 0, width = 0.45\columnwidth]{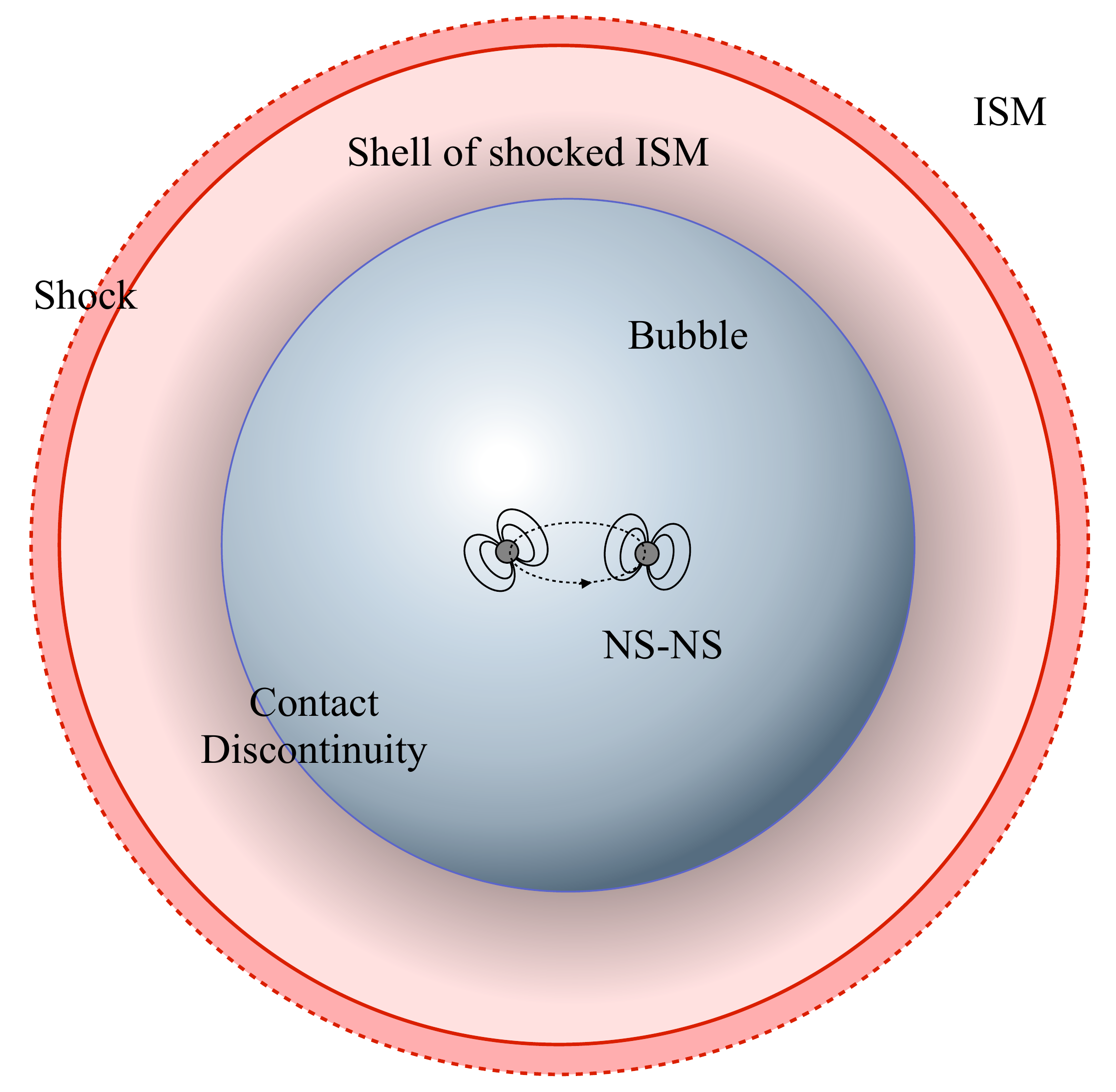}
\caption{Schematic representation of the bubble+shock system around a NS-NS binary.}
\label{system}
\end{figure}

The binary evolution and its orbital period are determined by the emission of gravitational waves. Besides, neutron stars possess large magnetic dipole moments, hence the electromagnetic energy is also extracted from the system via Poynting flux. Theoretical estimates and numerical modeling of NS mergers suggest significant amounts of energy, $\sim 10^{45}$~erg \citep{McW+L11,APiro12,E+12}, to be released in the electromagnetic form, and even more for magnetar binaries. The actual energy extraction mechanism may be rather complicated. Here we make an analogy with pulsar electrodynamics. A rapidly spinning neutron star magnetosphere produces a magnetized, relativistic electron-positron wind that exerts spin-down torque and extracts rotational energy \citep{GJ69}. The presence of relativistic leptonic plasma in the magnetosphere changes its structure drastically, which complicates analytical analysis. However, despite the structure of the force-free magnetosphere is different from the dipole field, the electromagnetic energy extraction rate given by the expression for the magnetic dipole radiation in vacuum \citep{GO69} is only within a factor of two, depending on geometry, of the exact rate obtained from direct numerical simulations \citep{S06}. Thus, for the case of a NS binary, we also expect that the electromagnetic energy is extracted in the form of the Poynting flux and it is safe to assume that the energy extraction rate  --- hereafter referred to as the {\em `Poynting luminosity'} or just the {\em `luminosity'}, $L(t)\equiv dE/dt$ --- is of the order of the electromagnetic losses in vacuum. Hence a bubble (or a cavity) forms in the surrounding medium. For simplicity, we assume spherical symmetry of the bubble and the uniform density of the ambient medium. These assumptions can readily be generalized to include environmental effects; for example, rapidly spinning NSs can themselves produce outflows and form pulsar wind nebulae around them. However, the uniform density assumption is reasonable if the neutron stars are non-rotating and do not have pulsar wind nebulae around them. The matter composition (relativistic plasma, magnetic field, waves) in the bubble depends on the exact mechanism of the energy extraction (e.g., a magnetized outflow, relativistic wind, electromagnetic and/or plasma wave emission and/or conversions, etc.), as well as the structure, evolution and interaction of magnetospheres of orbiting (and spinning) NSs --- the problem yet to be solved. However, regardless of the composition, the material in the bubble has a relativistic equation of state, $\gamma=4/3$. Because of the electromagnetic nature of the process producing the bubble and because the equation of state of matter inside it is that of electromagnetic radiation, we colloquially refer to them as the {\em `electromagnetic bubbles'}. Being a strong function of the orbital frequency, the binary Poynting luminosity $L(t)$ and, hence, the pressure inside the bubble are rapidly increasing around the merger time. The expanding bubble pushes onto the ambient medium at an increasing rate and can drive a shock in it, as is illustrated in Fig. \ref{system}. As the shock forms, it can be detected, e.g., via synchrotron radiation from accelerated electrons in shock-amplified magnetic fields.

In this paper we calculate the dynamics of the electromagnetic bubble and the bubble-driven shock around merging double neutron star or magnetar binaries and make observational predictions. In Section \ref{s:binary} we derive the evolution of the binary orbital period assuming the orbits are circular. In Section \ref{s:lumin} we discuss various processes of electromagnetic energy extraction and evaluate the Poynting luminosity. Section \ref{s:bubble} presents the model of the bubble+shock model and gives its evolution in the analytical form, including the finite-time-singular solution. Sections \ref{s:num} and \ref{s:obs} provide numerical estimates of the bubble+shock system parameters and make observational predictions, respectively. Section \ref{s:concl} presents discussion and conclusions.

\section{Binary inspiral}
\label{s:binary}

The dynamics and merger of a NS binary is determined by emission of gravitational waves. Here we assume circular orbits of the NSs, for simplicity. We also neglect general relativistic effects, though they will be important in the final few seconds of the merger. We also consider the NSs as point masses, hence neglecting tidal effects. With these assumptions, the energy loss rate in the systems of two gravitating bodies of masses $M_1$ and $M_2$ orbiting in circular orbits about their common center of mass is 
\beq
\frac{dE}{dt}=-\frac{G}{45c^5}\dddot Q^2=-\frac{32G}{5c^5}m^2r^4\Omega^6.
\label{gravrad}
\eeq
Here $m=M_1M_2/(M_1+M_2)$ is the reduced mass, $\Omega$ is the orbital frequency, $G$ is the gravitational constant, $Q_{\lambda\sigma}=m(3x_\lambda x_\sigma-r^2\delta_{\lambda\sigma})$ is the quadruple moment, ${\bf r}={\bf r_1}-{\bf r_2}$ is the mass separation vector with $r$ and $x_\lambda$ being its magnitude and components, $\dddot Q^2=\dddot Q^{\lambda\sigma}\dddot Q_{\lambda\sigma}^*$ and $\delta_{\lambda\sigma}$ is the Kronecker tensor. By the virial theorem $m v^2/2=K=-E=-U/2=GM_1M_2/2r=(m/2)\left[G(M_1+M_2)\Omega\right]^{2/3}$. NSs are relativistic objects so it is convenient to normalize the distance and the frequency by the radius of the maximally spinning Kerr black hole having the total mass of the system, $R_g$, and the Keplerian frequency at $R_g$:
\beq
R_g={G(M_1+M_2)}/{c^2}, \qquad \Omega_g=\left[{G(M_1+M_2)}/{R_g^3}\right]^{1/2}.
\label{Rg}
\eeq
Then 
\beq
r/R_g=(\Omega/\Omega_g)^{-2/3} ~\text{  and  }~ v/c=\left(\Omega/\Omega_g\right)^{1/3}. 
\label{ratios}
\eeq
In the $R_g$-units, the radius of a neutron star is $R_{NS}\simeq 3R_g$ (for equal mass stars) and the shortest separation distance in the binary (when NSs merge) is $R_m\sim 2 R_{NS}\sim6 R_g$. Realistically, disintegration of the neutron stars may occur at even larger separations due to strong tidal forces. Moreover, general relativistic effects become important at the final stage of the inspiral and merger, but they are omitted from consideration in our simple model. We parameterize the shortest separation distance with $\kappa_R$ as
\beq
R_m=\kappa_R R_g, \quad \kappa_R\simeq6.
\label{kappaR}
\eeq
Note that the smallest separation between the solar mass NS and BH in the NS-BH binary is smaller, $R_m\sim R_{NS}+R_{BH}\sim4 R_g$, i.e., $\kappa_R\simeq 4$.

From equation (\ref{gravrad}), noting that $R_g\Omega_g=c$ and $-E=K=(mc^2/2)(\Omega/\Omega_g)^{2/3}$, one has the equation for the orbital frequency
\beq
\frac{d(\Omega/\Omega_g)}{dt}=\frac{96}{5}\frac{M_1M_2}{(M_1+M_2)^2}\,\Omega_g\left(\frac{\Omega}{\Omega_g}\right)^{11/3}.
\eeq
It has a general solution with the {\em finite time singularity} 
\beq
\Omega(t)=\Omega_i(1-t/t_s)^{-3/8},
\label{Omega}
\eeq
where $\Omega_i$ is the initial orbital frequency and the source time is
\beq
t_s=\frac{5}{256}\frac{(M_1+M_2)^2}{M_1M_2}\left(\frac{\Omega_i}{\Omega_g}\right)^{-8/3} \Omega_g^{-1}.
\label{ts0}
\eeq
Note that because $r\gtrsim R_m\simeq6R_g$ there is maximum frequency and time beyond which equation (\ref{Omega}) is inapplicable,
\beq
\Omega_m/\Omega_g=\kappa_R^{-3/2},\quad t_m/t_s=1-\kappa_R^4(\Omega_i/\Omega_g)^{8/3}.
\label{tm}
\eeq

\section{Poynting luminosity $L(t)$}
\label{s:lumin}

A merging NS binary is a system of orbiting (and possibly spinning) magnetic dipoles. Because of strong magnetic fields in and rapid motion of the magnetospheres, the induced electric fields are strong enough to develop an electron-position cascade \citep{T10}, which loads the magnetospheres with leptonic relativistic plasma; hence the force-free configurations are formed and maintained. Therefore, the electromagnetic (EM) energy extraction can, in general, be different from the simple dipole/quadrupole radiation losses. Force-free modeling of magnetospheres of close NS-NS has not been done, so the details of the EM energy extraction remain unclear. However, we can use the analogy with pulsar electrodynamics, which indicates that the energy is extracted in the form of relativistic leptonic winds and field-aligned currents --- generally referred to as the Poynting flux ---  and, moreover, the EM losses are only a factor of two or less away from the low-frequency electromagnetic wave emission losses due to the magnetic dipole radiation in vacuum \citep{S06}, though some studies \citep{T06} suggest a more complicated picture. Therefore, we use the standard EM losses (dipole, quadrupole) in our estimates of the power taken away via the Poynting flux. One should keep in mind that these are estimates only; the actual mechanism and the electromagnetic energy extraction rate are unknown at present.

In general, the emitted power depends on the orientation of the dipole moment, the orbit eccentricities and the angular momentum of the system, as well as the spins of the neutron stars. Below, we consider the induced electric dipole and magnetic dipole radiation losses, and we briefly discuss quadrupole radiation losses. If the NSs are not rapidly spinning, the most efficient EM energy loss is the induced electric dipole emission, which occurs due to the orbital motion of magnetic moments of NSs. It is present even if the NSs are non-spinning at all. However, the electric dipole is of the order of $\sim v/c$ of the magnetic moment. Hence, the electric dipole emission contribution can be sub-dominant if the NSs are rapidly spinning. Moreover, it is natural to expect that during the final moments of binary inspiral, the NSs may be tidally locked and their angular frequency will be equal to the orbital angular velocity. In this regime, the magnetic dipole emission losses will dominate and the induced dipole emission will be $v^2/c^2$ times weaker. Quadrupole emission losses will also be present, but it is naturally weaker by a factor of $v^2/c^2$ and effectively renormalizes the induced electric dipole emission power.  

It will be shown that in all cases, the Poynting luminosity of the binary obeys the finite-time-singular (FTS) law, cf., Eq. (\ref{Omega}):
\beq
L(t)=L_s(1-t/t_s)^{-p},
\label{fts}
\eeq
where the index $p$ depends on the type of radiation mechanism, $L_s$ is the initial luminosity (at $t=0$) and $t_s$ is the binary lifetime -- the time until merger in the Newtonian approximation with the binary members being point-like objects. In reality, they are of finite size, so the actual merger time $t_m$ is less than $t_s$, as we discussed in the previous section. From equations (\ref{ts0}) and (\ref{tm}), one has
\beq
t_s-t_m=\frac{5\kappa_R^4}{256}\frac{(M_1+M_2)^2}{M_1M_2}\,\Omega_g^{-1}.
\label{tstm0}
\eeq
Note that this quantity is independent of initial conditions of the binary insipral, i.e., of the initial orbital frequency, $\Omega_i$. The total energy output does not diverge and is estimated to be
\bea
E_s &=& \int_0^{t_m}L_s(1-t/t_s)^{-p}\,dt
\nonumber\\
&=&(p-1)^{-1}L_s t_s\left[(1-t_m/t_s)^{-(p-1)}-1\right]
\nonumber\\
&\simeq& (p-1)^{-1}L_s t_s(1-t_m/t_s)^{-(p-1)},
\label{Es}
\eea
where the last approximate equality holds true if $p>1$ and $t_s-t_m\ll t_s$.

\subsection{Induced electric dipole losses}

Here we consider the simplest case of {\em non-spinning} NSs with magnetic moments $\boldsymbol{\mu}_1$ and $\boldsymbol{\mu}_2$ aligned with the orbital angular momentum. A magnetic moment moving non-relativistically with a velocity ${\bf v}$ induces an electric dipole moment ${\bf d}=({\bf v}_d\times \boldsymbol{\mu})/c$. For the NS system at hand, ${\bf d}_1=\hat{\bf r}_1\mu_1v_1/c$ and ${\bf d}_2=\hat{\bf r}_2\mu_2v_2/c$, where ${\bf v}_1=\dot{\bf r}_1 =\dot{\bf r} M_2/(M_1+M_2)$,\  ${\bf v}_2=\dot{\bf r}_2 =\dot{\bf r} M_1/(M_1+M_2)$ and ``hat'' denotes a unit vector.

The dipole radiation power is
\beq
\frac{dE_\text{EM}}{dt}=\frac{2}{3c^3}\ddot d^2=\frac{2}{3c^3}\frac{v^2}{c^2}\mu^2 \Omega^4,
\eeq
where we used that $d=(\mu v/c) e^{i\Omega t}$. 
The Poynting luminosity of the NS binary is the sum of the luminosities of each of the dipoles,
\beq
L(t)=\frac{2}{3c^3}\frac{(\mu_1^2M_2^2+\mu_2^2M_1^2)}{(M_1+M_2)^2}\left(\frac{\Omega}{\Omega_g}\right)^{2/3}\Omega^4\propto\Omega^{14/3},
\label{emrad}
\eeq
where we used Eq. (\ref{ratios}). With equation (\ref{Omega}), the luminosity becomes 
\beq
L(t)=L_s(1-t/t_s)^{-7/4},
\label{L1}
\eeq
where
\beq
L_s=\frac{2}{3c^3}\frac{(\mu_1^2M_2^2+\mu_2^2M_1^2)}{(M_1+M_2)^2}\,\Omega_g^4\left(\frac{\Omega_i}{\Omega_g}\right)^{14/3}.
\eeq
Using Eq. (\ref{tm}) and that $R_g\Omega_g=c$, one can estimate the peak luminosity $L_m=L(t_m)$ to be
\bea
L_m&=& L_s\kappa_R^{-7}\left(\frac{\Omega_i}{\Omega_g}\right)^{-14/3} 
\nonumber\\
&=& \frac{2}{3\kappa_R^7}\frac{\Omega_g}{R_g^3}\frac{(\mu_1^2M_2^2+\mu_2^2M_1^2)}{(M_1+M_2)^2}.
\label{Lm0}
\eea 
Note that this quantity is independent of the initial orbital frequency, $\Omega_i$. The total energy output is estimated to be
\beq
E_s \simeq \frac{5\kappa_R^4}{192}\frac{L_m}{\Omega_g}\frac{(M_1+M_2)^2}{M_1 M_2}.
\label{Es0}
\eeq
We note that in this case the Poynting luminosity index is $p=7/4$ in Eq. (\ref{fts}).

\subsection{Magnetic dipole losses}

We can also assume that the neutron stars are tidally locked and their magnetospheres rotate with the orbital frequency and result in the magnetic dipole losses with the power
\beq
\frac{dE_\text{EM}}{dt}=\frac{2}{3c^3}\ddot \mu^2=\frac{2}{3c^3}\mu^2 \Omega^4.
\label{emrad2}
\eeq
Using equation (\ref{Omega}), one obtains the luminosity of two tidally locked NSs
\beq
L(t)=L_s(1-t/t_s)^{-3/2},
\label{Lt1}
\eeq
where 
\beq
L_s=\frac{2}{3c^3}\left(\mu_1^2+\mu_2^2\right)\Omega_g^4\left(\frac{\Omega_i}{\Omega_g}\right)^{4} .
\eeq
The peak luminosity $L_m=L(t_m)$ is obtained using Eq. (\ref{tm}) to be
\bea
L_m&=& L_s\kappa_R^{-6}\left(\frac{\Omega_i}{\Omega_g}\right)^{-4} 
\nonumber\\
&=& \frac{2}{3\kappa_R^6}\frac{\Omega_g}{R_g^3}(\mu_1^2+\mu_2^2).
\label{Lm1}
\eea 
Finally, the total energy output is 
\beq
E_s \simeq \frac{5\kappa_R^4}{128}\frac{L_m}{\Omega_g}\frac{(M_1+M_2)^2}{M_1 M_2}.
\label{Es1}
\eeq
and we remind that the magnetic dipole losses result in the Poynting luminosity index is $p=3/2$ in Eq. (\ref{fts}).

\subsection{Magnetic quadrupole losses}

An orbiting magnetic dipole has a magnetic quadrupole moment $Q\sim \mu r$. Quadrupole losses are generally weak, but the induced electric dipole losses are, in fact, of the same order in $v/c$. Here we just look for a scaling. The emitted power is
\beq
\frac{dE_\text{EM}}{dt}=\frac{1}{45c^5}\dddot Q^2\propto r^2\Omega^6\propto\Omega^{14/3},
\eeq
which yields $p=7/4$  as in the induced electric dipole case. Thus, this energy loss channel just makes a correction to the induced electric dipole one.

\subsection{Comments on magnetar binaries, NS-BH binaries some NS-NS binaries}

Magnetars are neutron stars with the surface magnetic fields of the order of or larger than the Schwinger field  $B_S\simeq4.4\times 10^{14}\text{ G}$. All the estimates in previous sections remain true with the magnetic moment being about three orders of magnitude larger than for normal NSs, $\mu_\text{Mag}\sim10^{33}\text{ G cm}^3$. This increases the luminosity and the total energetics by a factor of one million, $E\propto\mu^2$ and hence $E\sim10^{47}\text{ erg}$, as is estimated in Section \ref{s:num}.

Evolution of a NS-BH binary is very similar to that of a NS-NS binary, except for the last moments before their merger when GR effects are important, hence equation (\ref{Omega}) holds. Poynting luminosity produced by a NS-BH binary is mostly via the electric dipole losses due to the electric dipole induced on a BH horizon, so one has (see Eq. (5) of \citealp{McW+L11})
\beq
L\propto v^2r^{-6}\propto\Omega^{14/3},
\eeq
which also yields $p=7/4$. However, if the NS companion is tidally locked, then the magnetic dipole losses would dominate the Poynting luminosity.

Finally, in NS-NS binaries, one of the members can have a substantially weaker magnetic field then another, as is the case in the double pulsar system PSR J0737-3039, for example. In this case, the stronger-field binary member induces the electric dipole moment on the other NS, very much like in the NS-BH binary. For such a case, a unipolar inductor model has been proposed \citep{APiro12}. This model generally predicts the Poynting flux luminosity to be similar to the case of the electric dipole losses, Eq (\ref{emrad}).

\section{Bubble+shock system evolution}
\label{s:bubble}

The formation and dynamics of bubbles and bubble-driven shocks has been studied in detail in a separate paper \citep{ML12}. Here we briefly outline the idea and use appropriate results. For simplicity, we assume that the electromagnetic bubble is spherical and expands into the interstellar medium (ISM) of constant density, as shown in Fig. \ref{system}, which is a good approximation for non-spinning neutron stars which do not have pulsar wind nebulae. The bubble is filled with matter with the relativistic equation of state parameterized by the adiabatic index $\gamma_\text{EM}=4/3$. The bubble surface acts as a piston and exerts pressure on the external medium producing an outgoing strong shock [the compression ratio is $\kappa=(\gamma+1)/(\gamma-1)\sim4$] in the cold unmagnetized ISM with mass density $\rho_\text{ISM}$ and the adiabatic index $\gamma=5/3$. The shell of shocked ISM is located in between the shock and the bubble. The mass density and pressure of the gas in the shell are determined by the Rankine-Hugoniot jump conditions. The pressure equilibrium throughout the system and across the bubble-shell interface (i.e., a contact discontinuity) is assumed. 

The time-dependent Poynting power $L(t)$ goes into the following components: the internal energies of the bubble, $dU_\text{bubble}$, and the shocked gas shell, $dU_\text{shell}$, the change of the kinetic energy of the bulk motion of the shell, $dK_\text{shell}$, assuming its swept-up mass $M_\text{swept}$ is constant, as well as heating, $dU_\text{@shock}$, and acceleration, $dK_\text{@shock}$, of the newly swept ISM gas at the shock. The $p\,dV$ work due to the expansion can be neglected because the external pressure is vanishing in the cold ISM. Thus, the master equation is 
\beq
L(t)\,dt=dU_\text{bubble} + dU_\text{shell} + \left.dK_\text{shell}\right|_{M_\text{swept}} 
+ dU_\text{@shock} + dK_\text{@shock}.
\label{master0}
\eeq
All the quantities can be expresses as a function of one dependent variable, the shock radius $R(t)$, for example. Other quantities follow straightforwardly, e.g., the shock velocity is $v=\dot R$, the bubble radius $R_b=(1-\kappa^{-1})^{1/3}R=(3/4)^{1/3}R$, and so on. The structure of the master equation is physically transparent: $L(t)\sim \dot K\sim d_t(\rho R^3 v^2)\sim \rho_\text{ISM}(c_1R^2\dot R^3+c_2R^3\dot R\ddot R)$, where $c_1$ and $c_2$ are some constants. The actual calculation \citep{ML12} yields: 
\beq
\frac{L(t)}{(4\pi/3)\rho_\text{ISM}} =c_1 R^2\dot R^3+c_2 R^3\dot R \ddot R,
\label{master}
\eeq
where
\bea
c_1&=&\frac{6}{(\gamma+1)^2}\left(\frac{\gamma_\text{EM}+1}{\gamma_\text{EM}-1}+2\right)
\simeq 7.6,
\\
c_2&=&\frac{4(\gamma_\text{EM}+1)}{(\gamma+1)^2(\gamma_\text{EM}-1)}+
\frac{12}{\gamma^2-1}\left[\left(\frac{\gamma+1}{2}\right)^{1/3}-1\right]
\simeq 4.6.
\eea
Solution of this inhomogeneous second-order nonlinear differential equation yields the shock radius as a function of time for any given luminosity law $L(t)$.

In Section \ref{s:lumin} we have shown that the Poynting luminosity of a NS binary is represented by the FTS law, Eq. (\ref{fts}),
\beq
L(t)=L_s(\Delta t/t_s)^{-p},
\eeq
where $\Delta t=t_s-t$. Approximate analytic solutions of Eq. (\ref{master}) exist at both early and late times \citep{ML12}. 
At early times, $t\ll t_s$, the binary has approximately constant Poynting luminosity, so the shock radius and velocity are:
\bea
R(t)&=&R_s\left({t}/{t_s}\right)^{3/5},
\label{Rss}\\
v(t)&=&v_s\left({t}/{t_s}\right)^{-2/5},
\label{Vss}
\eea
where 
\beq
R_s\sim\left(\frac{3}{4\pi}\frac{L_s t_s^3}{\rho_\text{ISM}}\right)^{1/5},\quad v_s\sim{R_s}/{t_s}
\label{Rs}
\eeq
and some numerical factors of order unity are suppressed for clarity. 
At late times, i.e., around the merger time $t\sim t_s$, the luminosity increases rapidly, so does the shock velocity, whereas the shock radius approaches a constant:
\bea
R(\Delta t)&=&R_s(\Delta t/t_s)^0,
\label{Rfts}\\
v(\Delta t)&=&v_s(\Delta t/t_s)^{-(p-1)/2}.
\label{Vfts}
\eea  
Note that the early-time solution is self-similar and the late-time one has a finite time singularity and therefore can break down if the velocity approaches the speed of light. Formally, it also breaks down at small $\Delta t\lesssim R/c$ because of the finite time needed for pressure equilibration throughout the system, where $c$ is the speed of light (recall, the material inside the bubble has a relativistic equation of state). Nevertheless, it can still be approximately true if $t_s$ is corrected for the finite light travel time, $t_s\to t_s+R/c$.

\section{Numerical estimates}
\label{s:num}

In previous sections we made theoretical estimates of the binary evolution, its Poynting luminosity and the evolution of the bubble+shock system. Here we make order of magnitude estimates.

First of all, if the masses of the compact companions are $M_1=M_2=m M_\sun$, then, from Eq. (\ref{Rg}) one has
\beq
R_g\sim3\times10^5\,m \text{ cm},\quad \Omega_g\sim10^5\,m^{-1}\text{ rad s}^{-1}.
\eeq
The binary lifetime, Eq. (\ref{ts0}), depends on the initial orbital angular speed, $\Omega_i$,
\beq
t_s\sim 2\times10^7\,m^{-5/3}\Omega_i^{-8/3}\text{ s}
\eeq
which is defined by the moment when $\Omega\to\infty$ or the separation of two point-like masses approaches zero. In nature, because of the finite size of the objects, the actual merger time $t_m$, Eq. (\ref{tstm0}), occurs before $t_s$: 
\beq
\Delta t_m\equiv t_s-t_m\sim 10^{-3}\,m\text{ s}.
\label{tm+}
\eeq
The maximum orbital frequency, Eq. (\ref{tm}), is
\beq
\Omega_m\sim 7\times10^{3}\,m^{-1}\text{ rad s}^{-1}.
\eeq
The orbital frequency evolution, Eq. (\ref{Omega}), can be re-written as
\bea
\Omega(\Delta t)&=&\Omega_m\left(\Delta t/\Delta t_m\right)^{-3/8}\\
&\sim&5\times10^2\,m^{-5/8}\Delta t^{-3/8}\text{ rad s}^{-1},
\eea
where $\Delta t\equiv t_s-t$ and other quantities are in CGS units unless stated otherwise.

Next, we estimate the Poynting luminosities and bubble+shock parameters for two cases, depending on the dominant mechanism of the electromagnetic energy extraction: the magnetic dipole and induced electric dipole losses. These two cases correspond to different physical scenarios. Case 1 represents the binary in which the NS spin periods are approximately equal to the orbital period, $\Omega_{NS}\sim\Omega$. \citet{BildstenCutler92} demonstrated that tidal locking is impossible in NS-BH binaries and is unlikely (though cannot be completely ruled out due to our ignorance about NS internal viscosity) in NS-NS binaries. However, complete locking, $\Omega_{NS}=\Omega$, is not required in this scenario. Partial synchronization will lower Poynting losses somewhat and can alter the temporal index, only. Rather strong partial synchronization, $(\Omega-\Omega_{NS})/\Omega\sim10\%$ has been estimated in \citep{APiro12} for the binary with orbital frequency $\Omega\sim10^2\ \text{s}^{-1}$ in which one of the NSs is non-magnetized. Hence Case 1 is entirely viable, and it is even more so given the lack of detailed understanding of interaction of NS magnetospheres. Case 2 corresponds to several physical scenarios: (i) the binary with slowly-spinning or non-spinning neutron stars, (ii) the NS-BH binary and (iii) the NS-NS binary with one of the NSs having much weaker magnetic field then the other. We note that Case 1 is more energetically efficient and effectively represents the upper bound on the process, whereas Case 2 is presumably more realistic.

\subsection{Case 1: magnetic dipole radiation losses} 

In this scenario, neutron stars are approximately synchronized, $\Omega_{NS}\sim\Omega$, hence the magnetic dipole losses dominate Poynting luminosity. Since this mechanism is the most energetically effective, Case 1 represents the order-of-magnitude upper limit on electromagnetic processes in the binary. The electromagnetic luminosity, Eq. (\ref{Lt1}), is
\beq
L(t)=L_s(1-t/t_s)^{-3/2}=L_m\left(\Delta t/\Delta t_m\right)^{-3/2}.
\label{Lt-1}
\eeq
From Eq. (\ref{Lm1}), the maximum luminosity, which occurs at the time of merger $L(t_m)$, is 
\beq
L_{m,NS}\sim 10^{44}\mu_{30}^2 m^{-4}\text{ erg s}^{-1}
\eeq
for a NS-NS binary and an order of magnitude larger for a NS-BH binary (because of lower $\kappa_R$). Here we assumed that the NS magnetic moments are $\mu_1=\mu_2=\mu$ and have a typical value of $\mu=\mu_{30} 10^{30}\text{ G cm}^3$. For a NS--magnetar or a double-magnetar binary, the luminosity is orders of magnitude larger:
\beq
L_{m,\text{Mag}}\sim 10^{50}\mu_{33}^2 m^{-4}\text{ erg s}^{-1}
\eeq
for a nominal value of the magnetic moment of $\mu_\text{Mag}\sim10^{33}\text{ G cm}^3$. 

The total EM energy produced in the process, Eq. (\ref{Es1}), is
\beq
E_{s,NS} \sim 2\times10^{41}\mu_{30}^2 m^{-3}\text{ erg},
\eeq
i.e., $E_{s,NS} \sim 10^{41}\text{ erg}$ for a typical NS binary, but can be as large as $E_{s,\text{Mag}}\sim10^{47}\text{ erg}$ for a magnetar binary. We stress that all calculations are done in the non-relativistic limit and do not account for general relativistic effects.

The normalization $L_s$ in Eq. (\ref{Lt-1}) depends on the binary lifetime, which is uncertain in reality. In fact, it is unreasonable to take the time since the binary was formed, because the Poynting luminosity is very low and other processes determine its ambient medium conditions. For example, motion of a binary in the ISM may disrupt and destroy the bubble by ram pressure if the bubble expansion velocity (at any moment of its evolution) is smaller than the bulk motion of the binary as a whole, which can be assumed to be a few tens km/s. As we will see below, this condition is satisfied if $t_s$ is about tens of years. For concreteness, we assume here $t_s\sim10\text{ years}$. One obtains
\bea
L_s &=& L_m t_s^{-3/2}(\Delta t_m)^{3/2}\nonumber\\
&\sim& 6\times10^{26}\ \mu_{30}^2 m^{-5/2} t_{s,10y}^{-3/2}\text{ erg s}^{-1},
\eea
where $t_{s,10y}\equiv t_s/(10\text{ yr})$.

The shock radius  and velocity evolve according to Eqs. (\ref{Rss}), (\ref{Vss}) at early times $t\ll t_s$ and according to Eqs. (\ref{Rfts}), (\ref{Vfts}) at later times around the merger time, $t\sim t_s$. The characteristic values are given by Eqs. (\ref{Rs}):
\bea
R_s &\sim& 10^{15}\,\mu_{30}^{2/5}m^{-1/2}n_{\text{ISM},0}^{-1/5}t_{s,10y}^{3/10}\text{ cm}, 
\label{Rs+}\\
v_s&\sim& 4\times 10^6\, \mu_{30}^{2/5}m^{-1/2}n_{\text{ISM},0}^{-1/5}t_{s,10y}^{-7/10}\text{ cm s}^{-1}, 
\label{Vs+}
\eea
that is, a typical size of the shock is $R_s\sim 70$~AU and its minimum velocity $v_s\sim40\text{ km/s}$ for the assumed parameters. At late times, the shock scalings with time are 
\beq
R(\Delta t)=R_s,\quad v(\Delta t)=v_s\left(\Delta t/t_s\right)^{-1/4}.
\label{RV}
\eeq

\subsection{Case 2: Induced electric dipole losses}

In this case the induced electric dipole (together with magnetic quadrupole) mechanism dominates, hence
\beq
L(t)=L_s(1-t/t_s)^{-7/4}=L_m\left(\Delta t/\Delta t_m\right)^{-7/4}.
\label{Lt-2}
\eeq
The maximum luminosity, Eq. (\ref{Lm0}), is 
\beq
L_{m,NS}\sim 5\times 10^{42}\mu_{30}^2 m^{-4}\text{ erg s}^{-1}
\eeq
for a NS-NS binary and about $5\times 10^{48}\text{ erg s}^{-1}$ for magnetars. The total energy, Eq. (\ref{Es0}), is
\beq
E_{s,NS} \sim 6\times10^{39}\mu_{30}^2 m^{-3}\text{ erg},
\eeq
i.e., $E_{s,NS} \sim 10^{40}\text{ erg}$ for a typical NS binary, but can reach $ \sim 10^{46}\text{ erg}$ for the magnetar case.

The time $t_s$ should be smaller in this case because of the lower overall energetics (see discussion in previous subsection). We use $t_s\sim3\text{ years}$. The normalization of the luminosity $L_s$ in Eq. (\ref{Lt-2}) is 
\bea
L_s &=& L_m t_s^{-7/4}(t_s-t_m)^{7/4}\nonumber\\
&\sim& 2\times10^{24}\ \mu_{30}^2 m^{-9/4} t_{s,1y}^{-7/4}\text{ erg s}^{-1}.
\eea
The characteristic values of the shock, Eqs. (\ref{Rs}), are
\bea
R_s &\sim& 10^{14}\mu_{30}^{2/5}m^{-9/20}n_{\text{ISM},0}^{-1/5}t_{s,1y}^{1/4}\text{ cm} \\
v_s&\sim& 3\times 10^6\, \mu_{30}^{2/5}m^{-1/2}n_{\text{ISM},0}^{-1/5}t_{s,1y}^{-7/10}\text{ cm s}^{-1},
\eea
i.e., the bubble and shock are an order of magnitude smaller in this scenario. At late times, the shock parameters scale as  
\beq
R(\Delta t)=R_s,\quad v(\Delta t)=v_s\left(\Delta t/t_s\right)^{-3/8}.
\eeq

\section{Observational predictions}
\label{s:obs}

Shocks can be observed via synchrotron radiation produced by shock-accelerated electrons in generated or amplified magnetic fields. We assume that the electrons and magnetic fields carry fractions $\epsilon_e$ and $\epsilon_B$ of the internal energy density, $u_\text{shell}\sim\rho_\text{ISM}v^2$, of the shocked gas\footnote{The exact calculation of $u_\text{shell}$ (Eq. 30 in  \citealp{ML12}), differs by a factor ${2}/{(\gamma^2-1)}=9/8$ which we suppress hereafter. }
\beq
\bar\beta_e\bar\gamma_em_ec^2n_{e,\text{shell}}=\epsilon_e u_\text{shell}, \qquad
{B^2}/{8\pi}=\epsilon_B u_\text{shell}, 
\eeq
where $\bar\beta_e\bar\gamma_e m_ec^2$ is the average energy of an electron, $\bar\beta_e=\bar v_e/c$ is the average dimensionless electron speed, and $n_{e,\text{shell}}=\kappa\,n_{e,\text{ISM}}\simeq \kappa\,n_\text{ISM}$ is the number density of electrons in the shocked gas shell, and $\kappa\sim4$ is the shock compression ratio. 

Here we consider only the first scenario with the magnetic dipole Poynting luminosity, since it provides the most interesting observations limits. From the observational point of view, the value of $t_s$ is nearly impossible to determine, whereas the bubble or shock radius can be measurable either directly (if the image is resolved) or indirectly (by time variability, for instance). Eliminating $t_s$ between Eqs. (\ref{Rs+}), (\ref{Vs+}) and using Eqs. (\ref{RV}), (\ref{tm+}) we can express the shock speed via the shock size:
\beq
v\sim 7\times10^8\, \mu_{31}m^{-5/4}n_{\text{ISM},0}^{-1/2}R_{s,15}^{-3/2}\Delta t_4^{-1/4}\text{ cm s}^{-1}, 
\label{VDt}
\eeq
where we evaluated the shock speed $10^4$ seconds before the merger and the neutron star surface field is $\sim10^{13}$ gauss. In estimates below we will use this expression with explicit dependence on the shock size. If desirable, the dependence on $t_s$ can readily be restored using Eq. (\ref{Rs+}).

The average Lorentz factor of accelerated electrons is obtained from
\bea
\bar\beta_e\bar\gamma_e(\Delta t) &\sim& \epsilon_e\kappa({m_p}/{m_e})\left({v}/{c}\right)^2
\nonumber\\
&\sim&4\ \epsilon_e\mu_{31}^2m^{-5/2}n_{\text{ISM},0}^{-1}R_{s,15}^{-3}\Delta t_4^{-1/2},
\eea
so the bulk of the electrons is mildly relativistic for a nominal $\epsilon_e\sim0.5$. The sub-equipartition magnetic field strength is
\bea
B(\Delta t) &\sim& \left(8\pi\epsilon_B m_p n_\text{ISM}\right)^{1/2}v
\nonumber\\
&\sim& 
5\times10^{-3}\ \epsilon_B^{1/2}\mu_{31}m^{-5/4}n_{\text{ISM},0}^{-1/2}R_{s,15}^{-3/2}\Delta t_4^{-1/4} ~\text{G},
\eea 
that is $B$ is of the order of 0.1 milligauss for a nominal $\epsilon_B\sim10^{-3}$ which means the field must be generated or amplified at the shock by an instability (such as Weibel, Bell, firehose, cyclotron), preexisting MHD turbulence or via other mechanism. 

Relativistic electrons emit synchrotron radiation with the peak of the spectrum being at the frequency 
\bea
\nu_s(\Delta t) &\sim& (2\pi)^{-1}\bar\gamma_e^2({eB}/{m_e c})
\nonumber\\
&\sim& 2\times10^5\ \epsilon_e^2\epsilon_B^{1/2} \mu_{31}^5m^{-25/4}n_{\text{ISM},0}^{-2}R_{s,15}^{-15/2}\Delta t_4^{-5/4}
~\text{Hz},
\eea
which is well below the self-absorption frequency (see \citealp{ML12}, for further details)
\bea
\nu_a&\sim&\left(10^{-2}\sigma_Tc\bar\gamma_en_\text{ISM}R_s/m_e\right)^{2/(s+4)}\nu_s^{(s-2)/(s+4)}
\nonumber\\
&\sim&10^8-10^9~\text{Hz}, 
\eea
where $\sigma_T$ is the Thompson cross-section and we assumed a power-law energy distribution of electrons with index $s$ with the nominal value of $\sim2.2$. Note, however, that the peak frequency, $\nu_s$, is a very strong function of the NS surface field and the bubble size. The peak frequency also depends on time and, formally, exceeds $10^9$~Hz at $\Delta t\la40$ seconds before the merger. 

Although the spectral peak is self-absorbed, we can still use $\nu_s$ to calculate the non-absorbed part of spectrum, since $P_\nu=P_{\nu,\text{max}}(\nu/\nu_s)^{-(s-1)/2}$ for the power-law distributed electrons. The spectral power at the peak (measured in erg~s$^{-1}$~Hz$^{-1}$) is $P_{\nu,\text{max}}(t)\approx P/\nu_s=(\sigma_T m_e c^2 /3e)B$, where $P$ is the total emitted power by a relativistic electron, $P=(4/3)\sigma_T c \bar\gamma_e^2(B^2/8\pi)$.

\begin{figure}[t]
\center
\includegraphics[angle = 0, width = 0.45\columnwidth]{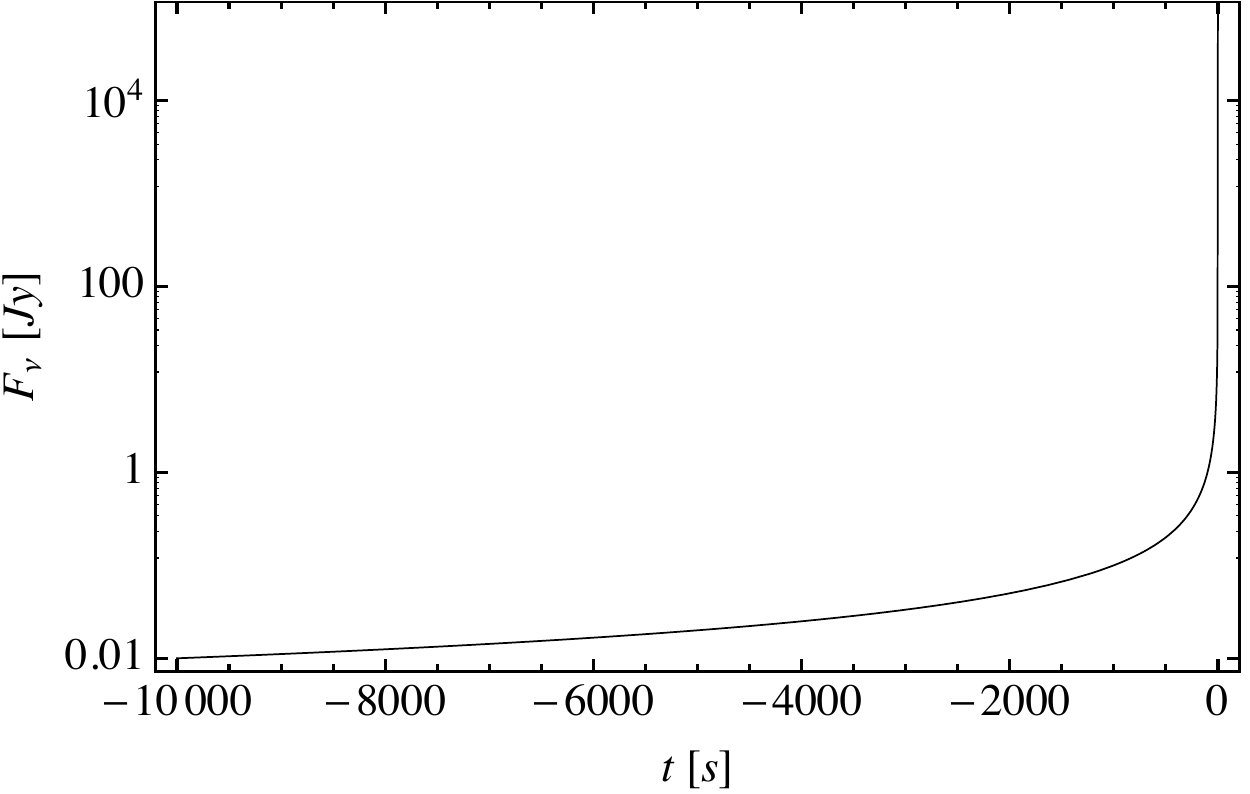}
\caption{Predicted spectral flux at $10^8$~Hz as a function of time. The NS-NS merger occurs at $t=0$, hence $\Delta t=-t$ in Eq. (\ref{Fnu-obs}).}
\label{fnu}
\end{figure}

The unabsorbed observed spectral peak flux from a source located in our galaxy at the distance $D=10^{22}~\text{cm}$ (i.e., $\sim 3$~kpc) would be $F_{\nu,\text{max}}=N_e P_{\nu,\text{max}}/(4\pi D^2)$, where $N_e=n_{e,\text{ISM}}V_\text{shock}$ is the total number of emitting electrons, hence
\bea
F_{\nu,\text{max}}(\Delta t) 
&\sim& \frac{1}{4\pi D^2}\left(\frac{4\pi}{3} R_\text{shock}^3 n_\text{ISM}\right)\frac{\sigma_T m_e c^2}{3e}B
\nonumber\\
&\sim& 0.6\ D_{22}^{-2}\epsilon_B^{1/2} \mu_{31}m^{-5/4}n_{\text{ISM},0}^{}R_{s,15}^{3/2}\Delta t_4^{-1/4}~\text{Jy}.
\eea
The spectrum above the self-absorption frequency scales as
\beq
F_\nu(\nu,t) \propto \nu^{-(s-1)/2}\Delta t^{-(3-5s)/8},
\label{Fnu-s}
\eeq
For the nominal value of $s=2.2$, we can estimate the observed flux at $\nu=10^8$~Hz as follows
\beq
F_{\nu}(\Delta t) 
\sim 0.01\ D_{22}^{-2}\epsilon_e^{1.2}\epsilon_B^{0.8} \mu_{31}^4m^{-5}n_{\text{ISM},0}^{-0.2}R_{s,15}^{-3}\nu_8^{-0.6}\Delta t_4^{-1}~\text{Jy}.
\label{Fnu-obs}
\eeq
This dependence is shown in Fig. \ref{fnu}.
Note that this flux is very sensitive to the masses of the binary members, the system size (larger bubbles are fainter) and the strength of the surface fields of NS (hence binaries with magnetars are much brighter, though $R_s$ for them is generally larger).

\section{Discussion and conclusions}
\label{s:concl}

In this paper we elucidated one aspect of the question of what happens around a merging binary of compact objects --- neutron stars, magnetars, black holes --- with either one or both binary members being magnetized (i.e., we did not consider a double black hole binary). The binary dynamics is known to be determined by the gravitational wave emission, which happens within a finite time, hence the orbital period has, formally, a finite time singularity (neglecting general relativistic, finite size, and other effects). Rapid orbital motion of the objects' magnetospheres also lead to substantial electromagnetic (or Poynting flux) losses amounting to $E_{EM}\sim10^{43}$ ergs for NSs with the surface magnetic field of $B\sim10^{13}$ gauss. This energy is extracted with relativistic, magnetized plasma, which forms an ``electromagnetic bubble'' of size $R_s\sim10^{15}$~cm. The bubble is expanding at an accelerated rate, as the binary approaches the merger time, and drives a shock in the ambient medium. For the finite-time-singular law of the Poynting luminosity, we have obtained the analytical solutions for the bubble+shock evolution as a function of time at both early and late times. Using these solutions we were able to address observational signatures of such systems. We have found that they can be observed as faint radio sources. The spectral flux from a source about 3~kpc away is estimated to be about a millijansky at $\nu\sim10^8$~Hz within $\Delta t\sim10^4$ seconds before the merger. The flux is increasing approximately as $\propto1/\Delta t$, so at $\Delta t\sim 100$ seconds before the merger, the flux is ~$\sim0.1$~Jy. The observed flux is a strong function of the NS magnetic field $F_\nu\propto B^4$ and the shock radius $F_\nu\propto R_s^{-3}$. The expected radio signal for magnetars is detectable by existing radio observatories. Due to the large total electromagnetic energy release $E_{EM}\sim10^{47}$~erg, the magnetar systems can be several tens times brighter than NS binaries. This estimate is, however, rather uncertain because of the unknown shock/bubble size, which is expected to be about an order of magnitude larger than for NS binaries. 

To make observational predictions, we assumed the Poynting luminosity index to be $p=3/2$. However, the actual value of $p$ depends on the mechanism of the electromagnetic energy extraction and may differ from 3/2. This index can be determined from observations via simultaneous monitoring of the flux as a function of time and frequency \citep{ML12}. If $F_{\nu}(t)\propto \nu^{\beta_\nu}(\Delta t)^{\beta_t}$, then the energy injection index is 
\beq
p=(1-5\beta_\nu-2\beta_t)/(1-5\beta_\nu). 
\label{p}
\eeq
Interestingly, the predicted radio emission light-curve differs from other radio emission signatures of neutron stars and their binaries, e.g., the pulsar wind nebula emission \citep{PiroKulkarni12} and the emission from the interaction of ejecta with the ISM \citep{NakarPiran11}. Particularly, our model predicts the very specific dependence of the flux on $\Delta t$, see Eq. (\ref{Fnu-s}), and the relation of the spectral and temporal indexes to the Poynting flux injection index, see Eq. (\ref{p}).

The scalings given in the paper are done for the shock co-moving time $\Delta t$, i.e., they do not include the finite light travel time from different parches of the spherical shock to the observer. When this effect is included, the radio signal from the shock of size $\sim10^{15}$~cm will be spread over time $\sim 2R_s/c$, which is several hours. Thus we make a prediction that a short GRB should be accompanied by radio emission, which starts a few hours before the main event, thus being a precursor, and lasts for several hours after it. The question of what happens after the merger, goes beyond the scope of this paper. We can speculate that since GRBs form narrow jets, the nearly spherical bubble will not be affected substantially, except within the jet. After the merger, the Poynting luminosity vanishes, so the bubble pressure drops due to expansion and the forward shock eventually loses pressure support and starts to expand freely. Not too long after the merger, when/if the magnetic energy is still large, the magnetically-driven GRB scenario may be realized \citep{LB03}. Long after the merger, the shock should settle onto the Sedov-von Neumann-Taylor solution of a point-source explosion with energy $E\sim10^{43}~\text{erg}$ (neglecting other possible energy inputs), thus resembling a ``micro supernova remnant". 

Several simplifying assumptions have been made in this study. In addition to the spherical symmetry, uniform external medium and neglect of general relativistic effects, there are a few others. Throughout the paper, we considered the shock to be strong, an assumption which may be wrong at early times (when the bubble expansion is slow because of low Poynting luminosity) and in relatively hot external medium, where the sound speed is comparable or larger then the bubble expansion speed --- in the latter case, the shock will not form at all. We have also assumed that the shock is non-relativistic. This is a good assumption for NS binaries, but may be violated in magnetar binaries at late times. We have also assumed pressure equilibrium throughout the system, so the finiteness of the light travel time (which is roughly the time of pressure equilibration) has not been taken into account. Numerical simulations are needed to address such time-dependent evolution. One can expect, for example, the formation of a reverse shock propagating into the bubble. We have also assumed that standard estimates of electromagnetic losses hold approximately true in the force-free magnetic configurations of orbiting companions. We based this assumption on numerical force-free simulations of pulsars, which generally confirm it. However, an accurate numerical modeling of both force-free structure and electron-positron cascades of orbiting magnetospheres is needed. Another interesting phenomenon here is that during the accelerated phase of the bubble+shock evolution, the bubble-shock interface can be prone to Rayleigh-Taylor instability, which can enhance mixing of the bubble material and the shocked ISM. Finally, we did not consider plasma dispersion effects which may alter the travel time for the radio signal compared to the prompt and afterglow emission at other energies.

We also note here on plasma processes at shock and details of particle acceleration, which impose additional constraints as follows. If the characteristic dynamical time of the system, which is $\sim\Delta t$, is longer than the inverse collisional frequency $\nu_\text{coll}^{-1}$ then a collisional shock forms. Otherwise,  when $\Delta t<\nu_\text{coll}^{-1}$, the shock is collisionless. In this case, if $\Delta t$ is (much) shorter then the Larmor frequency in the ambient field, the shock structure is not sensitive to the ISM field, but, instead, is determined by kinetic plasma instabilities (e.g., electrostatic Buneman or electromagnetic Weibel-like ones driven by particle anisotropy at the shock). The shortest associated timescale is the ion plasma time $\omega_{pi}^{-1}\sim10^3n^{1/2}~\text{s}$ and, moreover, it takes about a hundred $\omega_{pi}^{-1}$ seconds for an electrostatic shock [or $\omega_{pi}^{-1}(v/c)$ seconds for a Weibel shock] to form and respond to changing conditions; it takes even longer for particle Fermi acceleration. Thus, the synchrotron shock model should be used with great caution for $\Delta t$ as short as a fraction of a second or less.

\acknowledgements

One of the authors (MVM) is grateful to Sriharsha Pothapragada for useful suggestions. This work was
supported in part by DOE grant  DE-FG02-07ER54940 and NSF grant AST-1209665 (for MVM), and NSF grant AST-0907890 and NASA grants NNX08AL43G and NNA09DB30A (for AL).

\end{document}